\documentstyle[preprint,aps]{revtex}
\tightenlines

\begin{document}
\title{Bonn potential and shell-model calculations for $N=126$ isotones}
\author{L. Coraggio$^{1,2}$, A. Covello$^1$, 
A. Gargano$^1$, N. Itaco$^1$, and T. T. S. Kuo$^{2}$}
\address{$^1$Dipartimento di Scienze Fisiche, Universit\`a
di Napoli Federico II, \\ and Istituto Nazionale di Fisica Nucleare, \\
Complesso Universitario di Monte S. Angelo, Via Cintia, I-80126 Napoli, Italy \\
$^2$Department of Physics, SUNY, Stony Brook, New York 11794}
\date{\today}

\maketitle
\begin{abstract}
We have performed shell-model calculations for the $N=126$ isotones $^{210}$Po,
$^{211}$At, and $^{212}$Rn using a realistic effective interaction derived 
from the Bonn A nucleon-nucleon potential by means of a $G$-matrix 
folded-diagram method. The calculated binding energies, energy spectra and 
electromagnetic properties show remarkably good agreement with the 
experimental data. The results of this paper complement those of our previous
study on neutron hole Pb isotopes, confirming that realistic effective 
interactions are now able to reproduce with quantitative accuracy the 
spectroscopic properties of complex nuclei.

\end{abstract}
\draft
\pacs{PACS number(s): 21.60.Cs; 21.30.Fe; 27.80.+w}

\section{Introduction}

During the past few years, we have studied a number of nuclei around doubly 
magic $^{100}$Sn and $^{132}$Sn \cite{andr96,andr97,cove98,andr99,covel99} 
in terms of the
shell model employing realistic effective interactions derived 
from the meson-theoretic Bonn A nucleon-nucleon ($NN$) potential 
\cite{mach87}. In these studies we have considered nuclei with few valence
particles or holes, their properties being of special interest
for a stringent test of the basic ingredients of shell-model calculations. 
The  aim of our work is to assess the ability of 
realistic effective interactions to provide a quantitative description of 
nuclear structure properties. This is in fact a key
point to understand if the time has come to make the shell model a truly
microscopic theory of nuclear structure.

As is well known, the first step in this direction was taken more than
thirty years ago by Kuo and Brown \cite{kuo66} who derived an $sd$-shell 
effective  interaction from the Hamada-Johnston potential 
\cite{hamada62}. 
Later on, an effective interaction for the lead region was 
derived \cite{herling72} by Kuo and Herling (KH) from the same potential. 
Since then, however, substantial progress has been made in both
the development of high-quality $NN$ potentials and the
many-body methods for calculating the matrix elements of the effective 
interaction.

As regards the first point, modern potentials reproduce quite accurately all
the known $NN$ scattering data. A review of recent 
developments in the
field of $NN$ potentials is given in Refs. \cite{mach94,mach99}. We
only recall here that two potentials which fit equally well the $NN$
data up to the inelastic threshold may differ substantially in their
off-shell behavior. Thus, different $NN$ potentials may produce
somewhat different nuclear structure results. 

As for the second  point,
an accurate calculation of the Brueckner $G$ matrix is now
feasible while the so-called folded-diagram expansion yields a rigorous 
expression for the model-space effective interaction $V_{\rm eff}$. The main 
aspects of the above derivation of $V_{\rm eff}$ are described in Refs. 
\cite{jiang92,kuo96}.

Based on these improvements, a new generation of realistic effective
interactions has become available, fostering renewed interest in realistic
shell-model calculations. It is in this context that our recent
studies of medium-mass nuclei are framed.
The remarkably
good agreement between theory and experiment obtained for these nuclei has
challenged us to perform the same kind of realistic shell-model 
calculations for heavy-mass nuclei. 
In a previous work \cite{cora98} we considered 
the neutron hole isotopes $^{206,205,204}$Pb.
Here, we present the results of a companion study of the $N=126$ isotones,
focusing attention on $^{210}$Po, $^{211}$At, and 
$^{212}$Rn. These nuclei, with two to four protons in the $Z=82-126$ shell, 
offer the opportunity to further test our realistic effective interactions in 
the lead region. 

The $N=126$ nuclei, as well as the lead isotopes, have 
long been the subject of both experimental and theoretical studies. From the
experimental point of view, these stable or near-stable nuclei have been 
extensively
investigated and a rather large amount of experimental data is available for them.
On the other hand, the good doubly magic character of $^{208}$Pb
has motivated many shell-model calculations in this region.
In all the calculations performed so far, however, 
phenomenological interactions have been used \cite{ma72,glaudemans85}, the 
only notable exception 
being the pioneering work  of McGrory and Kuo \cite{mcgrory75}, where 
the KH interaction  was
employed.

The outline of the paper is as follows. In Sec. II we give an outline of
our calculations, including a brief review of the derivation of the
effective interaction. In Sec. III we present the results obtained for
binding energies, energy spectra and electromagnetic properties,
comparing them with the experimental data.
Sec. IV contains a discussion and a summary of our conclusions.

\section{Outline of Calculations}
We assume that $^{208}$Pb is a closed core and let the valence
protons occupy the six single-particle (sp) orbits $0h_{9/2}$,
$1f_{7/2}$, $0i_{13/2}$, $1f_{5/2}$, $2p_{3/2}$, and $2p_{1/2}$.
 We  take the sp energies from the
experimental spectrum of $^{209}$Bi \cite{nndc}.
They are (in MeV): $\epsilon_{h_{9/2}}=0.0$,
$\epsilon_{f_{7/2}}=0.896$, $\epsilon_{i_{13/2}}=1.609$,
$\epsilon_{f_{5/2}}=2.826$, $\epsilon_{p_{3/2}}=3.119$,
$\epsilon_{p_{1/2}}=3.633$.

As already mentioned in the Introduction, we use in the present
calculation a realistic effective
interaction derived from the Bonn A free $NN$ potential.
Let us now outline our derivation of $V_{\rm eff}$.
Because of the strong repulsive core contained in the Bonn A potential, which
is a feature common to all modern $NN$ potentials, the model-space $G$
matrix corresponding to the chosen $V_{NN}$ must be calculated first.
The $G$ matrix is defined \cite{krenc76}  by the integral equation:

\begin{equation}
G(\omega)=V+VQ_2 \frac{1}{\omega-Q_2TQ_2}Q_2G(\omega),
\end{equation}

\noindent
where $V$ represents the $NN$ potential, $T$ denotes the two-nucleon kinetic
energy, and $\omega$ is an energy variable (the so-called starting energy).
The two-body Pauli exclusion operator $Q_2$ prevents double counting, namely
the intermediate states allowed for $G$ must be outside of the chosen model
space.
Thus the Pauli operator $Q_2$ is dependent on the model space, and
 so is the  $G$ matrix.
The operator $Q_2$ is specified, as discussed in Ref. \cite{krenc76}, by a set
of three numbers $(n_1,n_2,n_3)$ each representing a shell-model orbital
(we number the orbits starting from
the bottom of the oscillator well; for instance, the orbit $0d_{5/2}$
is denoted as orbit 4 and $0h_{11/2}$ as orbit 16).
Note that in Eq.(1) the Pauli exclusion operator $Q_2$ is defined in
terms 
of harmonic oscillator wave functions 
while plane-wave functions are employed for the intermediate states
of the $G$ matrix.  

Since the valence-proton and -neutron orbits outside $^{208}$Pb are different,
 our $Q_2$ operators for protons and for neutrons
are different and consequently our $G$-matrix calculation is considerably
more complicated than in the case when the two operators are the same.
In the present calculation we have fixed $(n_1,n_2,n_3)=(22,45,78)$ 
for the neutron orbits, and
$(n_1,n_2,n_3)=(16,36,78)$ for the proton orbits.
Our procedure for calculating the $G$ matrix is outlined below. We first
 calculate the free $G$ matrix $G_F$ in a proton-neutron
representation, $G_F$ being defined by
\begin{equation}
G_F=V+V\frac{1}{e}G_F,
\end{equation}
with $e\equiv (\omega - T)$. Note that $G_F$ does not contain the 
Pauli exclusion operator and hence its calculation is relatively convenient.
 Then we calculate the Pauli correction term 
\cite{krenc76,tsai72},
\begin{equation}
\Delta G=-G_F\frac{1}{e}P_2\frac{1}{P_2(\frac{1}{e}+\frac{1}{e}G_F\frac{1}{e}
)P_2}P_2
\frac{1}{e}G_F,
\end{equation}
where $P_2 = 1 - Q_2$, separately for protons and for neutrons. 
Finally the full $G$ matrix as defined
by Eq.(1) is obtained as
\begin{equation}
G=G_F +\Delta G.
\end{equation}

For the harmonic oscillator parameter $\hbar \omega$ we use the value
6.88 MeV, as obtained from the expression $\hbar \omega=45A^{-1/3}-25A^{-2/3}$
for $A=208$.
Note that in the present work we have chosen the value of $n_2$ so 
as to include three
harmonic oscillator shells above the Fermi level denoted by $n_1$.
In earlier works on light- and medium-mass nuclei, $n_2$ was fixed by taking
into account only two major shells above the $n_1$-th orbit.
For instance, a common choice for $sd$-shell calculations is $n_2$=10 with
$n_1$=3 \cite{jiang92}.
For the $N=126$ isotones, however, the present choice is more
appropriate.
In fact, these calculations as well as those for 
lead isotopes \cite{cora98,cove99}
have led to a substantially better agreement with experiment
when in the derivation of the  effective interaction $n_2$ has been increased
from two to three shells above the $n_{1}$-th orbit.

Using the above $G$ matrix we then calculate the irreducible vertex
function $\hat{Q}$-box, which is composed of irreducible valence-linked
$G$-matrix diagrams through second order in $G$.
These are precisely the seven first- and second-order diagrams considered by
Shurpin {\em et al.} \cite{shurp83}.
The effective interaction can be written in operator form as

\begin{equation}
V_{eff} = \hat{Q} - \hat{Q'} \int \hat{Q} + \hat{Q'} \int \hat{Q} \int
\hat{Q} - \hat{Q'} \int \hat{Q} \int \hat{Q} \int \hat{Q} + ~...~~,
\end{equation}

\noindent
where $\hat{Q}$ is the $\hat{Q}$-box, and the integral sign represents a
generalized folding operation \cite{kuo80}.
$\hat{Q'}$ is obtained from $\hat{Q}$ by removing terms of first order in the
reaction matrix $G$.
After the $\hat{Q}$-box is calculated, the energy-independent $V_{\rm eff}$ is
then obtained by summing up the folded-diagram series of Eq.(5) to all orders
using the Lee-Suzuki iteration method \cite{suzuki80}.
This last step can be performed in an essentially exact way for a given
$\hat{Q}$-box.

Once the effective interaction has been derived, the shell-model
calculations are carried out employing the OXBASH code \cite{brown}.

As regards the electromagnetic observables, we have calculated them
by making use of effective operators \cite{mav66,krenc75} which 
take into account core-polarization effects.
More precisely, by using a diagrammatic description as in Ref. \cite{mav66},
we have only included first-order diagrams in $G$. 
This implies that folded-diagram renormalizations are not necessary
\cite{krenc75}. 

As is well known, the nuclear magnetic properties may be significantly
affected by mesonic exchange currents.
An estimate of their contribution for nuclei in the vicinity of
$^{208}$Pb has been given in Refs. \cite{hyuga73,towner77}.
This amounts to renormalizing the gyromagnetic factor $g_l$ from the
bare value of $g_l = 1$ to $1.155$ and $g_s$ from $5.586$ to $5.699$.
We have made use of these values in our calculation of the effective
$M1$ operator. 

\section{Results and comparison with experiment}

In this section the results of our calculations for the three nuclei 
$^{210}$Po,
$^{211}$At, and $^{212}$Rn are presented and compared with experiment. 
The energy spectra are presented separately for each isotone in the three
following subsections. For $^{210}$Po a detailed comparison between 
calculated and observed spectroscopic factors is also reported. The last
subsection is devoted to the discussion of the electromagnetic properties
of all three nuclei. 

The calculated ground-state binding energies relative to $^{208}$Pb are 
compared with the observed values \cite{audi93} in Table I. The mass 
excess value for
$^{209}$Bi needed for absolute scaling of the sp levels was taken from 
\cite{audi93}.
As regards the Coulomb interaction between the valence
protons, we have assumed that it gives a contribution proportional to 
the number of interacting
proton pairs, namely $E_C = \frac{n(n-1)}{2} V_C$ ($n$ is the number
of valence protons). The strength $V_C$ has been taken to be 270 keV,
which is the value of the matrix element of the Coulomb force between 
two $h_{9/2}$ protons with $J=0$.

From Table I we see that
a very good agreement with experiment is obtained for all three nuclei. In
fact, the calculated binding energy for $^{210}$Po falls practically within the
error bar, while the other two calculated values differ by less
than 100 keV from the experimental ones. 

\subsection{Spectrum of $^{210}$Po}

The experimental \cite{nndc} and theoretical spectra of 
$^{210}$Po are compared in 
Fig. 1. Here all the calculated and experimental levels up to 3.3 MeV are
reported, while in the higher-energy region the negative-parity states have
been excluded (in the experimental spectrum we have also omitted three states
without spin and parity assignment). 
The calculated states which are shown in 
Fig. 1 are all the 44 states arising from the configurations $h_{9/2}^2$, 
$h_{9/2}f_{7/2}$, $h_{9/2}i_{13/2}$, $f_{7/2}^2$, $i_{13/2}^2$, 
$h_{9/2}f_{5/2}$,
and $h_{9/2}p_{3/2}$. In the energy region 3.5-4.5 MeV we find the eight
negative-parity states of the $f_{7/2}i_{13/2}$ configuration, but for
them, as mentioned above, we have not tried to establish any correspondence
with observed levels. In fact, in this energy interval negative-parity states
have been observed which cannot be described within our model space.
As an example, we mention the $12^-$ and $13^-$ states, which cannot be
constructed in our model space, and  the three $5^-$ levels observed at
3.43, 3.70, and 3.71 MeV to which corresponds only one calculated $5^-$ state
at 3.85 MeV. These experimental ``extra'' levels arise from core
excitations, and in some cases significant
admixtures of these excitations and two-particle model-space states are
likely to occur.

In Fig. 1 we see that the calculated spectrum is characterized
by four groups of levels: the first one up to 1.6 MeV, the second between
2.2 and 2.5 MeV,  the third between 2.7 and 3.3 MeV, and the fourth above
3.8 MeV. The five levels of the first
group are dominated by the $h_{9/2}^2$ configuration, while the second group
contains all the members of the $h_{9/2}f_{7/2}$ multiplet, besides the $0^+$
state arising from the $f_{7/2}^2$ configuration. The other three states of
this latter configuration together with all the states arising from the
$h_{9/2}i_{13/2}$ configuration are in the third group. All the states
in these three groups, with few exceptions, are almost pure, the percentage 
of the
dominant configuration ranging from 95 to 100\%. Only for the ground 
state and the $0^{+}_{2}$ and $2^{+}_{3}$ states the contribution coming from
configurations other than the dominant one is particularly significant, 
the percentage of such 
configurations being 21, 28, and 16\%, respectively. The 17 levels arising
from the configurations $i_{13/2}^2$, $h_{9/2}f_{5/2}$, and 
$h_{9/2}p_{3/2}$ all lie in the energy interval 3.8-4.8 MeV. It should be
mentioned, however, that for most of these states the wave functions are not
quite pure. In particular, we find that a significant admixture of the
three above configurations is present in the even $J$ states. 

Up to 3.3 MeV excitation energy each state of a given $J^\pi$ in the 
calculated spectrum can be unambiguously associated with an observed level,
the only exception being the $2^{+}_{3}$
state at 2.95 MeV excitation energy. However, two levels with no
angular momentum and 
parity assignment have been observed  at 2.66 and 2.87 MeV, and in Ref. 
\cite{mann88} it
is suggested that the 2.87-keV $\gamma$ ray measured   in the 
$^{209}$Bi$(t,2n)^{210}$Po reaction is a good candidate for the 
$2^{+}_3 \rightarrow 0^{+}_{gs}$ transition. The experimental 
$3^{-}_{1}$ state at
2.39 MeV, as well as the the $5^{-}_{1}$ and $4^{-}_{2}$ states at 2.91 and
3.11 MeV, respectively, have no theoretical counterpart. In fact,
the first one reflects the collective nature of the octupole $3^-$ state at
2.61 MeV in $^{208}$Pb, while the other two levels arise from the neutron
particle-hole configuration $\nu(g_{9/2}p_{1/2}^{-1})$ \cite{mann88}, 
and therefore cannot be described within our model space. It should be
noted that each of the two above $5^{-}_{1}$ and $4^{-}_{2}$ levels 
lies very close in energy to 
the state with the same $J^{\pi}$ originating from the $h_{9/2}i_{13/2}$
configuration. Therefore it cannot be excluded, as we shall see when discussing
the spectroscopic factors, that some mixing occurs between single-particle and
core-excited states. 

Above 3.8 MeV only 10 out of the 17 levels arising from the configurations 
$i_{13/2}^2$, $h_{9/2}f_{5/2}$, and 
$h_{9/2}p_{3/2}$ have been experimentally identified. For all of them,
except the $(7^{+},4^{+})$ state at 4.55 MeV, a correspondence with 
states predicted by the
theory can be safely established. As for the $(7^{+},4^{+})$ state, it may be
associated with either the $7^{+}_{2}$ or the $4^{+}_{5}$ calculated states, 
which lie at 4.48 and 4.52 MeV excitation energy, respectively. In a very recent
work \cite{klein99} the assignment $7^+$ has been proposed for the 
experimental level at 4.55 
MeV and a new level with $J^{\pi}=(4^{+})$ has been identified at 4.54 MeV.

As regards the quantitative agreement, we see from Fig. 1 that it is very
satisfactory, the discrepancies between calculated and experimental
excitation energies being less than 100 keV for most of the
states. More precisely, including the level at 2.87 MeV (identified as
a $J^{\pi} = 2^+$ state) as well as the $J^{\pi}=4^{+}$ and
$7^+$ of Ref. \cite{klein99}, 37 observed levels
have been associated with states predicted by the theory, and only for seven of
them the experimental and calculated excitation energies differ by more than 
100 keV.  The rms deviation $\sigma$ \cite{sigma} relative to these 
37 levels is 87 keV. 

In Ref. \cite{groleau80} the $^{209}$Bi$(^{3}$He,$d)^{210}$Po and
$^{209}$Bi$(^{4}$He,$t)^{210}$Po reactions have been studied and the 
single-proton strengths of transitions to various
excited levels in $^{210}$Po have been  extracted from the
measured cross sections. These observed strengths
are compared with the calculated values in Table II, where we also list the 
experimental and theoretical excitation energies.
The experimental uncertainties appearing in Table II are
only statistical and the numbers in parentheses, which correspond to levels
not fully resolved, were extracted by a peak fitting procedure 
(see Ref. \cite{groleau80}).
The theoretical spectroscopic factor $S$ is defined as

   $$ S^{J^{\pi} \beta}_{l_{j}} =\frac {1}{2J+1} 
         |\langle ^{210}{\rm{Po}}, \, J^{\pi},\, \beta \parallel 
         a_{lj}^{\dagger}
         \| ^{209}{\rm{Bi}}, \, J^{\pi}_i=9/2^-, \, gs \rangle|^{2},$$

\noindent{where we assume that the ground state of $^{209}$Bi is a single 
$h_{9/2}$ proton outside the doubly magic $^{208}$Pb. The label
$\beta$ specifies states of $^{210}$Po with the same $J^{\pi}$.}

From Table II we see that for almost all the states  
of the three low-lying multiplets the agreement between theory and experiment
is  very good. Actually, a significant discrepancy exists only for the 
$11^-$ and the second $5^-$ state. However,
it was suggested in Ref. \cite{groleau80}  that the level at 2.85 MeV 
was an unresolved
doublet with $J^{\pi}=11^{-}$ and $3^-$, as it has been found later
to be the case \cite{nndc}.
Thus the observed strength, 3.25, attributed to the $11^-$ state has to 
be compared to the sum  of the two calculated  strengths relative to the 
$11^-$ and $3^-$ states, which are 2.30 and 0.69, respectively. As for 
the $5^-$ state, our calculation overestimates the experimental value.
Part of the  single particle strength is contained in
the first $5^-$ state, which, as mentioned above, is not predicted by the 
theory, being essentially a core-excited state. 

In the region above 3.8 MeV the values of the measured strengths 
($l=1$ and 3) are 
generally smaller than those relative to the states of the three low-lying
multiplets. On the other hand, as it was already pointed out 
at the beginning of
this section, the calculated wave functions of 
several states in this region show a strong
admixture of the configurations $i_{13/2}^2$, $h_{9/2}f_{5/2}$, and 
$h_{9/2}p_{3/2}$. 
Thus, a comparison between theory and experiment may provide a test of 
the calculated percentages of the
$h_{9/2}f_{5/2}$ and $h_{9/2}p_{3/2}$ configurations (obviously, the 
contribution of the $i_{13/2}^{2}$ configuration is not determined
directly from the 
measured strengths). It should also be noted
that the experimental data do not allow to distinguish between $p_{3/2}$ and
$p_{1/2}$ transfers. We have found, however, that a small component of the
$h_{9/2}p_{1/2}$ configuration is present only in the $4^{+}_{4}$ and 
$4^{+}_{5}$ states. From Table II we see to the observed
strengths of the states above 3.8 MeV are
quite well reproduced by the theory. Note that the level at 4.55 MeV
excited via $f_{5/2}$ transfer is likely to correspond to an unresolved
doublet with $J^{\pi}=4^{+}$ and $7^{+}$ (see discussion above).
In this case the measured strength, 1.83, should be 
interpreted as the sum of the calculated strengths 1.50 and 0.20 
relative to the
$7^{+}_{2}$ and $4^{+}_{5}$ states, respectively.
In this connection, it should be pointed out that the observed
strength of the level at 4.55 MeV excited via $l = 1$ transfer and
assigned $J^{\pi} = 4^+$ \cite{groleau80} is also well reproduced 
by our calculation.

\subsection{Spectrum of $^{211}$At}

The experimental \cite{nndc} and theoretical spectra of $^{211}$At
are compared in Fig. 2, where all the observed 
levels up to 3.3 MeV excitation energy are reported. In the calculated spectrum
all the levels up to about 2.0 MeV are included while
in the higher-energy region only the states which can be associated to the
observed ones are reported.  For the sake of completeness all the calculated
excitation energies up to 2.7 MeV are listed in  Table III.

From Fig. 2 we see that  a one-to-one correspondence can
be established between the experimental and calculated levels up to 1.5 MeV,
the only exception being  the experimental 
$(\frac{9}{2},\frac{11}{2},\frac{13}{2})$ level at 1.35 MeV which can be 
associated to  either  the $(\frac{9}{2}^{-})_{2}$ or $(\frac{13}{2}^+)_1$ 
calculated state.
As regards the two observed 
levels with no firm spin assignment at 1.12 and 1.23 MeV, we propose
the assignment $J^{\pi} = \frac{11}{2}^-$ and $ \frac{15}{2}^-$, respectively. 

Above 1.5 MeV many more levels than the
observed ones are predicted by our calculations. In particular, in the energy interval 1.5-2.0 MeV we find
12 levels, only three of them having the experimental counterpart. 
These three states have  
$J^{\pi} = (\frac{3}{2})^-$, $(\frac{23}{2}^-)$, and $ (\frac{5}{2})^-$
and can be identified with the calculated states with the same 
angular momentum and parity at 1.82, 1.97, and 2.04 MeV, respectively.
It should be noted that we predict the
existence of a lower-lying $\frac{5}{2}^-$ state at 1.85 MeV.

As regards the states above 2.0 MeV, we identify the 
experimental $J^{\pi}=\frac{1}{2}^-, \frac{3}{2}^-$ level at 2.06 MeV with the
calculated one with $J^{\pi}=\frac{1}{2}^-$ at 2.11 MeV, thus
confirming the tentative assignment of Ref. \cite{astner72}. As far as 
the lowest $\frac{1}{2}^+$ state is concerned, 
the calculated energy is 2.78 MeV, 
namely 300 keV higher than that of the  observed one. However, this
state, which was  populated  in a first-forbidden $\beta^+$ decay of 
the ground state of $^{211}$Rn \cite{astner72}, has been interpreted as 
a core-excited $^{212}$Rn $\otimes (\pi s_{1/2})^{-1}$ state and, 
as such, is not expected to be adequately 
reproduced within  our model space. In Ref. \cite{astner72} the nature
and assignment of the  $J=\frac{1}{2},\frac{3}{2}$ level at 2.65 MeV was also
discussed and it was suggested that it originates from a mixing of the two 
$^{212}$Rn $\otimes (\pi d_{3/2})^{-1}$ and $^{211}$At(gs) 
$\otimes ^{210}$Pb$(3^{-})$
core-excited states. 
We may only mention here that up to 3.5 MeV our calculations predict 
the existence of the three $\frac{3}{2}^+$ states reported in Fig. 2
and of no other $\frac{1}{2}^+$ aside the above
mentioned one.

The quantitative agreement between our results and
experiment is very satisfactory. In fact, the discrepancies 
for the excitation energies are all in the order of few tens of keV, the 
only exception being the $J^{\pi} =\frac{29}{2} ^+$ state, 
which comes about 170 keV below its experimental counterpart. Excluding 
the $\frac{1}{2}^+$ state and the two levels at 1.35 and 2.65 MeV, 
for which we have not attempted any identification, the $\sigma$ 
value is only 64 keV.

It should be noted that
all the ten levels arising from the $h_{9/2}^3$ configuration lie at an 
excitation energy smaller than 1.5 MeV. In this energy region we also
find the two seniority $v=1$
states of the $h_{9/2}^2 f_{7/2}$ and $h_{9/2}^{2} i_{13/2}$ 
configurations.
The latter states as well as the ground state contain, however,   
significant configuration mixing. In fact,  the  percentage of configurations
other than the dominant one is 22\% in the ground and 
$(\frac{7}{2}^{-})_{1}$ states, reducing to 14\% in the 
$(\frac{13}{2}^{+})_{1}$ state. In all other levels up to 1.5 MeV
the percentage of the dominant configuration ranges from 90 to 98\%. 
Above 1.5 MeV  the
negative-parity states are members of the multiplet
$h_{9/2}^{2} f_{7/2}$, while all the positive-parity ones originate from the
$h_{9/2}^{2} i_{13/2}$ configuration, except the $(\frac{27}{2}^{+})_2$
state, which arises from the $h_{9/2} f_{7/2} i_{13/2}$ configuration.
All these states are essentially pure, the only exception being the
$(\frac{9}{2}^-)_4$ state, which contains 38\% of the $h_{9/2}
f_{7/2}^2$ configuration.

\subsection{Spectrum of $^{212}$Rn}

Rather little experimental information \cite{nndc} is presently 
available for $^{212}$Rn.
Up to about 4 MeV only 22 excited states have
been identified (nine of them with unknown spin and parity), while
our calculations predict a much higher level density. In particular,
in the low-energy region (up to 2.6 MeV) we find 27 states compared
to 8 in the experimental spectrum.
In this situation, any attempt to associate calculated states with 
experimental 
levels without assigned spin and parity may be misleading. 
Therefore, in Fig. 3 we exclude such states in the experimental spectrum
and report only  those yrast and yrare calculated states which
are candidates for the observed levels. For completeness, all the calculated
excitation energies up to about 2.6 MeV are listed in Table IV. 
It should be mentioned that above 4 MeV excitation energy
several high-spin states have been observed. In Fig. 3,
however, we do not include these levels, since 
their description is likely to require that core-excited states be
explicitly taken into account.

From Fig. 3 we see that the calculated spectrum reproduces very well the
experimental one, the discrepancies being
smaller than 100 keV  for the energies of 9 out of the 13 states
considered.
The rms deviation $\sigma$ is only 85 keV, in line with the
values obtained for the two lighter isotones. 

As regards the structure of the states, we find that the wave
functions of the seven higher-lying levels are
substantially pure.
These states are members of the three multiplets $h_{9/2}^{4}$ 
($J^{\pi}=10^{+},12^{+}_{1}$), $h_{9/2}^{3}f_{7/2}$ 
($J^{\pi}=12^{+}_{2},14^{+}$), and $h_{9/2}^{3}i_{13/2}$ 
($J^{\pi}=15^{-},16^{-},17^{-}$), and the percentage
of the dominant configuration is at least 95\%. This is not the case for the
lower-lying states, whose wave functions contain significant configuration
mixing. In the first four excited states the percentage of the dominant 
configuration, $h_{9/2}^{4}$, ranges from 71 to 76\% while it becomes 55\%
in the ground state. 
As for the $8_2^+$ and $11^-$ states, the percentages of the
dominant configurations, $h_{9/2}^{3}f_{7/2}$ and
$h_{9/2}^{2}i_{13/2}$, are 80 and 82\%, respectively.

\subsection{Electromagnetic properties}

The effective operators needed for the calculation of electromagnetic
observables have been derived as described in Sec. II.

In Table V we compare the experimental magnetic moments in $^{210}$Po,
$^{211}$At and $^{212}$Rn \cite{nndc,ragha89} with the calculated values. 
We see that the agreement is remarkably good in all cases.
Only two $M1$ reduced
transition probabilities are known  in $^{211}$At \cite{nndc,bayer95}.
They are compared
with our theoretical results  in Table VII. We see that both the 
calculated and experimental values are extremely small.

Let us now come to the electric observables. In Tables VI and VII we compare
the calculated quadrupole moments and $E2$, $E3$ transition rates 
with the experimental ones \cite{nndc,ragha89,bayer95,stuchbery93,becker91}. 
The agreement is very good, the only discrepancy
regarding the $B(E2;2^+_1 \rightarrow 0^+_1)$ in $^{210}$Po.
It should be mentioned, however, that the experimental value was
obtained by comparing the cross section $\sigma(2^{+},\,
^{210}{\rm Po})$ measured in a $^{210}$Po study by inelastic scattering 
of deuterons  with the corresponding one
for $^{206}$Pb \cite{ellegaard73}. 
As far as the quadrupole moments are concerned, 4 out of the 5
calculated values are within the error bars and the observed signs, when 
measured, are correctly reproduced.
It is worth noting that our results do not differ significantly from
those obtained using an effective proton charge $e_p^{\rm eff}=1.5e$, 
which is consistent with the values adopted by other authors 
\cite{glaudemans85,mcgrory75}.

As regards the $B(E3)$'s, they are all underestimated by our calculations.
It is well known that enhanced $E3$ transitions 
in nuclei in the lead region can be taken as a signature of mixing 
of the $3^-$ core excitation into
the involved levels \cite{bergstrom85}.
We should note, however, that whereas our calculations fail to reproduce
the $B(E3)$ values, a good description of the states involved
in such transitions is obtained for the excitation energies 
as well as for the other
electromagnetic properties. Thus, these states are likely to
contain very small components  of octupole excitation which, however, 
are sufficient to largely enhance the $E3$ transition rates.
In particular, the $E3$ transitions in $^{211}$At and $^{212}$Rn,
and the $11_1^- \rightarrow 8_2^+$ transition in $^{210}$Po
correspond to the single-particle transition $i_{13/2} \rightarrow
f_{7/2}$, which
is expected to be very fast owing to the coupling
between the $f_{7/2}$ orbital and the $3^-$ collective state \cite{bohr75}.
The $11_1^- \rightarrow 8_1^+$ transition in $^{210}$Po is instead
of the type $i_{13/2} \rightarrow h_{9/2}$ and is slowed down by spin
flip.

\section{Discussion and conclusions}

In this work, we have performed shell-model calculations for the $N=126$
isotones $^{210}$Po, $^{211}$At, and $^{212}$Rn, employing an effective
interaction derived from the Bonn A nucleon-nucleon potential by means of a
$G$-matrix folded-diagram method. As for the single-proton energies, we
have taken them from the experimental spectrum of $^{209}$Bi.
It should be stressed that, since we have also derived in a microscopic way
the effective operators needed for the calculation of electromagnetic
observables, no use has been made of any adjustable parameter.

These calculations, as well as the previous ones on neutron hole Pb 
isotopes \cite{cora98,cove99}, are the first in the $^{208}$Pb 
region where a modern
realistic interaction has been used. As already mentioned in the
Introduction, the first attempt to employ in this region effective interactions derived
from the free nucleon-nucleon potential dates back to the
early 70s \cite{mcgrory75}. In that work, however, the Hamada-Johnston $NN$
potential was used and only the 3p-1h core-polarization diagram (the
so-called bubble) was
included in the calculation of the effective interaction. It should
also be noted that to obtain good agreement with experiment
for the $^{204-206}$Pb isotopes, the bubble was multiplied 
by the empirical factor 0.75.

As regards our calculations, we have obtained a very good description
of both $N=126$ isotones and Pb isotopes in a truly microscopic
way. It cannot be said, however, that our agreement with experiment is
much better than that obtained in Ref. \cite{mcgrory75}.  
In this connection, we
found it worthwhile to calculate the complete energy spectrum of $^{210}$Po
up to 5 MeV
making use of the KH effective interaction. It turned out that the
$\sigma$ value relative to the 37 levels considered in Sec. III A is
116 keV, namely only about 30 keV larger than our value. On the other
hand, it should be mentioned that a comparison between the results
of the two calculations evidences more substantial differences.
More precisely, the calculation with the KH interaction predicts some
levels to lie 300-400 keV above those obtained with our effective
interaction. We do not feel, however, that a detailed comparison
between the two kinds of calculations is very meaningful.
We consider as the main achievement of our studies of nuclei around
$^{208}$Pb to have shown that effective interactions derived from the
Bonn A potential by means of a $G$-matrix folded-diagram approach
lead to a quite accurate description of these nuclei.
This outcome acquires more relevance when considered along with the
results of our studies on nuclei with few valence particles or
holes in the region of doubly magic $^{100}$Sn and
$^{132}$Sn \cite{andr96,andr97,cove98,covel99}.
In fact, the remarkable overall agreement with experiment obtained in
all cases considered leads to the conclusion that the new
generation of realistic effective interactions is quite adequate for
nuclear structure calculations.

Actually, being focused on identical particle systems, our studies provide
a stringent test of the isospin $T=1$ matrix elements of the effective 
interaction. A test of the $T=0$ matrix elements is of course equally 
important. In this connection, it may be mentioned that in earlier
works \cite{jiang92} it turned out that not enough attraction was provided
by the calculated matrix elements of the $T=0$ effective interaction, 
which has a stronger dependence on the tensor force strength than the 
$T=1$ interaction. We should point out, however, that in a recent 
study  \cite{andr99} of the doubly odd nucleus $^{132}$Sb we have obtained 
results which are as good as those regarding like nucleon systems.
Along the same lines we are currently studying other nuclei with both neutrons 
and protons outside closed shells.

A main question relevant to microscopic nuclear structure calculations is
the extent to which they depend on the $NN$ potential used as input. We
are currently trying to explore this problem. Preliminary calculations
indicate that different $NN$ potentials produce somewhat different nuclear 
structure results \cite{andr96,gat99}. In particular, it has turned
out 
that the best agreement with experiment is produced by the Bonn A potential.

In conclusion, at the present stage of our investigation of the role of realistic effective 
interactions in complex nuclei, it is our belief that a truly
microscopic description of nuclear structure properties is now within reach.

\acknowledgements
\noindent
{This work was supported in part by the Italian Ministero dell'Universit\`a 
e della Ricerca Scientifica e Tecnologica (MURST) and by the U.S. Grant
No. DE-FG02-88ER40388.}

\begin{figure}
\caption{Experimental and calculated spectrum of $^{210}$Po.}
\end{figure}

\begin{figure}
\caption{Experimental and calculated spectrum of $^{211}$At.}
\end{figure}

\begin{figure}
\caption{Experimental and calculated spectrum of $^{212}$Rn.}
\end{figure}

\newpage

\mediumtext
\begin{table}
\setdec 0.00
\caption{Experimental and calculated ground-state binding energies (MeV) 
relative to $^{208}$Pb for $^{210}$Po, $^{211}$At, and $^{212}$Rn.}

\begin{tabular}{lrr}
Nucleus & \multicolumn {2} {c} {Binding energy}  \\ 
& Expt.~~~~~ & Calc. \\
\tableline
$^{210}$Po & $8.782 \pm 0.004$ &8.789    \\
$^{211}$At & $11.765 \pm 0.005$ &11.816    \\
$^{212}$Rn & $16.065 \pm 0.006$ &16.146    \\
\end{tabular}
\end{table}

\newpage

\mediumtext
\begin{table}
\setdec 0.00
\caption{Comparison of the experimentally observed spectroscopic strengths
from the $^{209}$Bi($\alpha,t$)$^{210}$Po and $^{209}$Bi($^{3}$He,$d$)
$^{210}$Po
reactions with the calculated values. See text for comments.}

\begin{tabular}{llllccc}
$l_{j}$ & $J^{\pi}$ & ${\rm{E_{exp}(MeV)}}$ & ${\rm{E_{calc}(MeV)}}$ &   
$\frac{(2J+1)}{(2J_{i}+1)}S(\alpha,t)$ &
$\frac{(2J+1)}{(2J_{i}+1)}S(^{3}{\rm He},d)$ &  Calc. \\
\tableline
$h_{9/2}$ & $2^{+}$ &1.181 &1.130& $1.00 \pm 0.06$  & $1.16 \pm 0.22$ & 0.98  \\
& $4^{+}$ & 1.427 &1.395&   $1.79 \pm 0.07$  & $1.58 \pm 0.28$ & 1.78  \\
& $6^{+}$ & 1.473 &1.493&   $2.64 \pm0 .09$  & $2.63 \pm 0.25$ & 2.58  \\
& $8^{+}$ & 1.557 &1.555&   $3.40 \pm0 .11$  & $3.42 \pm 0.26$ & 3.36  \\
$f_{7/2}$ & $8^{+}$ &2.188&2.179& $1.71 \pm0 .03$  & $1.64 \pm 0.05$ & 1.68  \\
&$2^{+}$ & 2.290 & 2.292&  $0.42 \pm0 .01$  & $0.42 \pm 0.02$ & 0.47  \\
&$6^{+}$ & 2.326 & 2.367&  $1.31 \pm0 .03$  & $1.26 \pm 0.04$ & 1.28  \\
&$4^{+}$ & 2.382 & 2.394&  (0.90)  & (0.90) & 0.88  \\
&$1^{+}$ & 2.393 & 2.220&  (0.31)  & (0.35) & 0.30  \\
&$5^{+}$ & 2.403 & 2.422&  (1.10)  & (1.10) & 1.10  \\
&$3^{+}$ & 2.414 & 2.380&  (0.72)  & (0.75) & 0.70  \\
&$7^{+}$ & 2.438 & 2.437&  $1.51 \pm 0.03$  & $1.50 \pm 0.06$ & 1.50  \\
$ i_{13/2}$ & $3^{-}$ & 2.846 &2.862&    &  & 0.69  \\
&$11^{-}$ & 2.849 &2.700&  $3.25 \pm0 .06$  &  & 2.30  \\
&$5^{-}$ & 2.910 &&  $0.31 \pm0 .01$  &  &   \\
&$9^{-}$ &  2.999 &3.016&  (1.88)  &  & 1.89  \\
&$7^{-}$ &  3.016 &3.065&  (1.53)  &  & 1.50  \\
&$2^{-}$ &  3.024 &2.682&  (0.54)  &  & 0.50  \\
&$5^{-}$ &  3.026 &3.024&  (0.78)  &  & 1.10  \\
&$4^{-}$ &  3.075 &3.039&  $0.78 \pm0 .02$  &  & 0.90  \\
&$6^{-}$ &  3.125 &3.097&  (1.34)  &  & 1.30  \\
&$8^{-}$ &  3.168 &3.121&  (1.66)  &  & 1.70  \\
&$10^{-}$ & 3.183 &3.154&  $2.11 \pm0 .04$  &  & 2.10  \\
$f_{5/2}$ & $2^{+}$ & 3.792 &3.828&  $0.35 \pm 0.02$   &  & 0.34  \\
&$4^{+}$ & 4.027 & 4.152&  0.60   &  & 0.67  \\
&$6^{+}$ & 4.139 & 4.256&  0.82   &  & 0.86  \\
&$3^{+}$ & 4.320 & 4.309&  $0.86 \pm 0.04$   &  & 0.69  \\
&$5^{+}$ & 4.382 & 4.391&  $1.19 \pm 0.05$   &  & 1.10  \\
&$6^{+}$ & 4.469 & 4.503&  $0.55 \pm 0.03$   &  & 0.43  \\
&$7^{+},4^{+}$ & 4.553 & 4.384,4.552&  $1.83 \pm 0.07$   &  & 1.50,0.20  \\
$p_{3/2}+p_{1/2}$ &$4^{+}$ & 4.027 &4.152&  & 0.03 & 0.07  \\
&$6^{+}$ & 4.139 & 4.256&     & 0.20 & 0.29  \\
&$6^{+}$ & 4.469 & 4.503&     & $0.56 \pm 0.04$ & 0.66  \\
&$4^{+}$ & 4.553 & 4.522&     & $0.35 \pm 0.07$ & 0.22  \\
&$3^{+}$ & 4.591 & 4.605&     & (0.75) & 0.69  \\
&$5^{+}$ & 4.624 & 4.673&     & (1.35) & 1.10  \\
&$6^{+}$ & 4.644 & 4.691&     & (0.55) & 0.31  \\
\end{tabular}
\end{table}

\newpage

\mediumtext
\begin{table}
\setdec 0.00
\caption{Calculated low-energy levels of $^{211}$At.}

\begin{tabular}{llllll}
$J^{\pi}$ & E(MeV) &$J^{\pi}$& E(MeV)&$J^{\pi}$& E(MeV)  \\ 
\tableline
$\frac{9}{2}^{-}$ & 0.0 & $\frac{5}{2}^{-}$ & 2.040& $\frac{19}{2}^{-}$&2.350  \\
$\frac{7}{2}^{-}$ & 0.679 & $\frac{7}{2}^{-}$ & 2.042& $\frac{9}{2}^{+}$&2.393  \\
$\frac{7}{2}^{-}$ & 0.783 & $\frac{11}{2}^{-}$ & 2.042& $\frac{11}{2}^{+}$&2.412\\ 
$\frac{5}{2}^{-}$ & 0.955 & $\frac{17}{2}^{-}$ & 2.045& $\frac{19}{2}^{+}$&2.427\\
$\frac{13}{2}^{-}$ & 1.053 & $\frac{9}{2}^{-}$ & 2.080& $\frac{29}{2}^{+}$&2.466\\
$\frac{11}{2}^{-}$ & 1.098 & $\frac{11}{2}^{-}$ & 2.110& $\frac{21}{2}^{+}$&2.472\\
$\frac{3}{2}^{-}$ & 1.103 & $\frac{1}{2}^{-}$ & 2.110& $\frac{25}{2}^{+}$&2.515\\
$\frac{9}{2}^{-}$ & 1.186 & $\frac{15}{2}^{+}$ & 2.124& $\frac{23}{2}^{+}$&2.528\\
$\frac{13}{2}^{+}$ & 1.236 & $\frac{3}{2}^{-}$ & 2.131& $\frac{11}{2}^{+}$&2.530\\
$\frac{15}{2}^{-}$ & 1.337 & $\frac{13}{2}^{+}$ & 2.136& $\frac{7}{2}^{+}$&2.539\\
$\frac{17}{2}^{-}$ & 1.339 & $\frac{21}{2}^{-}$ & 2.178& $\frac{3}{2}^{+}$&2.541\\
$\frac{21}{2}^{-}$ & 1.467 & $\frac{7}{2}^{-}$ & 2.181& $\frac{5}{2}^{+}$&2.549\\
$\frac{9}{2}^{-}$ & 1.631 & $\frac{5}{2}^{-}$ & 2.185& $\frac{11}{2}^{-}$&2.560\\
$\frac{7}{2}^{-}$ & 1.681 & $\frac{9}{2}^{-}$ & 2.189& $\frac{9}{2}^{+}$&2.574\\
$\frac{11}{2}^{-}$ & 1.721 & $\frac{13}{2}^{-}$ & 2.204& $\frac{13}{2}^{+}$&2.589\\
$\frac{3}{2}^{-}$ & 1.824 & $\frac{9}{2}^{-}$ & 2.229& $\frac{27}{2}^{+}$&2.609\\
$\frac{5}{2}^{-}$ & 1.856 & $\frac{11}{2}^{-}$ & 2.238& $\frac{5}{2}^{-}$&2.617\\
$\frac{13}{2}^{-}$ & 1.929 & $\frac{15}{2}^{-}$ & 2.245& $\frac{13}{2}^{-}$&2.625\\
$\frac{15}{2}^{-}$ & 1.940 & $\frac{17}{2}^{+}$ & 2.279& $\frac{17}{2}^{+}$&2.629\\
$\frac{9}{2}^{-}$ & 1.967 & $\frac{13}{2}^{-}$ & 2.282& $\frac{9}{2}^{-}$&2.686\\
$\frac{23}{2}^{-}$ & 1.969 & $\frac{17}{2}^{-}$ & 2.289& $\frac{9}{2}^{+}$&2.688\\
$\frac{19}{2}^{-}$ & 1.987 & $\frac{15}{2}^{-}$ & 2.291& & \\
\end{tabular}
\end{table}

\newpage

\mediumtext
\begin{table}
\setdec 0.00
\caption{Calculated low-energy levels of $^{212}$Rn.}

\begin{tabular}{llll}
$J^{\pi}$ & E(MeV) &$J^{\pi}$& E(MeV)  \\ 
\tableline
$0^{+}$ & 0.0 & $3^{+}$ & 2.276  \\
$2^{+}$ & 1.221 &$4^{+}$ & 2.277    \\
$4^{+}$ & 1.506 &$5^{+}$ & 2.290     \\
$6^{+}$ & 1.619 &$5^{+}$ & 2.357    \\
$8^{+}$ & 1.677 &$7^{+}$ & 2.387   \\
$4^{+}$ & 2.057 &$4^{+}$ & 2.450    \\
$8^{+}$ & 2.122 &$3^{+}$ & 2.473    \\
$0^{+}$ & 2.170 &$0^{+}$ & 2.499    \\
$6^{+}$ & 2.177 &$6^{+}$ & 2.561    \\
$2^{+}$ & 2.198 &$2^{+}$ & 2.581    \\
$1^{+}$ & 2.208 &$11^{-}$ & 2.597    \\
$2^{+}$ & 2.211 &$8^{+}$ & 2.632    \\
$6^{+}$ & 2.226 &$3^{-}$ & 2.651     \\
$7^{+}$ & 2.274 &$10^{+}$ & 2.655    \\
\end{tabular}
\end{table}

\newpage

\mediumtext
\begin{table}
\setdec 0.00
\caption{Calculated and experimental dipole moments (in nm).}
\begin{tabular}{llcc}
 Nucleus & $J^{\pi}$ & \multicolumn{2}{c} {$\mu$} \\
 ~~~&~~~& Calc. & Expt. \\
\tableline 
 $^{210}$Po & $6^+_1$               & +5.29  & $\pm5.48 \pm 0.05$ \\
 ~~~~~~~~~  & $8^+_1$               & +7.06  & $+7.35 \pm 0.05$ \\
 ~~~~~~~~~  & $11^-_1$              & +13.12 & $+12.20 \pm 0.09$ \\
 $^{211}$At & $(\frac{15}{2}^-)_1$  & +6.6   & $\pm6.8 \pm 0.6$ \\
 ~~~~~~~~~~ & $(\frac{21}{2}^-)_1$  & +9.32  & $+9.56 \pm 0.09$ \\
 ~~~~~~~~~~ & $(\frac{29}{2}^+)_1$  & +16.23 & $+15.31 \pm 0.13$ \\
 $^{212}$Rn & $4^+_1$               & +3.56  & $\pm4.04 \pm 0.24$ \\
 ~~~~~~~~~~ & $6^+_1$               & +5.308 & $\pm5.454 \pm 0.048$ \\
 ~~~~~~~~~~ & $8^+_1$               & +7.064 & $+7.152 \pm 0.016$ \\
 ~~~~~~~~~~ & $14^+_1$              & +15.07 & $\pm14.98 \pm 0.42$ \\
 ~~~~~~~~~  & $17^-_1$              & +18.45 & $\pm17.85 \pm 0.17$ \\
\end{tabular}
\end{table}

\mediumtext
\begin{table}
\setdec 0.00
\caption{Calculated and experimental quadrupole moments (in $e$b).}
\begin{tabular}{llcc}
 Nucleus & $J^{\pi}$ & \multicolumn{2}{c} {Q} \\
 ~~~&~~~& Calc. & Expt. \\
\tableline 
 $^{210}$Po & $8^+_1$               & -0.588 & $-0.552 \pm 0.020$ \\
 ~~~~~~~~~  & $11^-_1$              & -0.92  & $-0.86 \pm 0.11$ \\
 $^{211}$At & $(\frac{21}{2}^-)_1$  & -0.54  & $\pm0.53 \pm 0.05$ \\
 ~~~~~~~~~~ & $(\frac{29}{2}^+)_1$  & -1.07  & $\pm1.01 \pm 0.19$ \\
 $^{212}$Rn & $8^+_1$               & -0.29  & $(-)0.17 \pm 0.02$ \\
\end{tabular}
\end{table}

\newpage

\mediumtext
\begin{table}
\setdec 0.00
\caption{Calculated and experimental reduced transition probabilities (in 
W.u.).}

\begin{tabular}{lllcc}
 Nucleus & Transition &$J^{\pi}_i \rightarrow J^{\pi}_f$ & \multicolumn{2}{c}
{Reduced transition probabilities (in W.u.)} \\
 ~~~&~~~&~~~& Calc.  & Expt. \\
\tableline
 $^{210}$Po & E2 & $2_1^+ \rightarrow  0_1^+$    & 3.55  & 
$0.56 \pm 0.12$ \\			      
 ~~~~~~~~~  & E2 & $4_1^+ \rightarrow  2_1^+$    & 4.46  & 
$4.53 \pm 0.15$ \\
 ~~~~~~~~~  & E2 & $6_1^+ \rightarrow  4_1^+$   & 3.07  & 
$3.00 \pm 0.12$ \\			      
 ~~~~~~~~~  & E2 & $8_1^+ \rightarrow  6_1^+$   & 1.25  &
$1.10 \pm 0.05$ \\			      
 ~~~~~~~~~  & E3 & $11_1^- \rightarrow 8_2^+$   & 6.1   &
$19.7 \pm 1.1 $ \\			      
 ~~~~~~~~~  & E3 & $11_1^- \rightarrow 8_1^+$   & 0.55  & 
$3.71 \pm 0.10$ \\

 $^{211}$At & E2 & $(\frac{3}{2}^-)_1 \rightarrow (\frac{5}{2}^-_1)$ 
& 10.1 & $12.5 \pm 1.4$ \\
 ~~~~~~~~~~ & E2 & $(\frac{3}{2}^-)_1 \rightarrow (\frac{7}{2}^-)_2$ 
& 1.67 & $1.77 \pm 0.17$ \\
 ~~~~~~~~~~ & E2 & $(\frac{3}{2}^-)_1 \rightarrow (\frac{7}{2}^-)_1$ 
& 0.15 & $0.39 \pm 0.04$ \\
 ~~~~~~~~~~ & E2 & $(\frac{15}{2}^-)_1 \rightarrow (\frac{11}{2}^-)_1$
& 2.3  & $1.3 \pm 0.3$ \\
 ~~~~~~~~~~ & E2 & $(\frac{21}{2}^-)_1 \rightarrow (\frac{17}{2}^-)_1$
& 2.60 & $2.51 \pm 0.05$ \\
 ~~~~~~~~~~ & E2 & $(\frac{29}{2}^+)_1 \rightarrow (\frac{25}{2}^+)_1$
& 1.6  & $1.8 \pm 0.5$ \\
 ~~~~~~~~~~ & M1 & $(\frac{3}{2}^-)_1 \rightarrow (\frac{5}{2}^-)_1$ 
& $8 \times 10^{-7}$ & $1.4 \times 10^{-4} \pm 0.4 \times 10^{-4}$ \\
 ~~~~~~~~~~ & M1 & $(\frac{15}{2}^-)_1 \rightarrow (\frac{13}{2}^-)_1$
& $8 \times 10^{-7}$ & $0.7 \times 10^{-4} \pm 0.2 \times 10^{-4}$ \\
 ~~~~~~~~~~ & E3 & $(\frac{29}{2}^+)_1 \rightarrow (\frac{23}{2}^-)_1$
& 6.3  & $20.1 \pm 1.8 $ \\
 $^{212}$Rn & E2 & $4_1^+ \rightarrow  2_1^+$  & 1.42  & 
$1.04 \pm 0.04$ \\			      
 ~~~~~~~~~  & E2 & $6_1^+ \rightarrow  4_1^+$  & 0.73  & 
$0.40 \pm 0.05$ \\			      
 ~~~~~~~~~  & E2 & $8_1^+ \rightarrow  6_1^+$  & 0.252 & 
$0.115\pm 0.006$ \\			      
 ~~~~~~~~~  & E2 & $12_1^+ \rightarrow 10_1^+$ & 3.6   & 
$4.4 \pm 0.2 $ \\			      
 ~~~~~~~~~  & E2 & $14_1^+ \rightarrow 12_1^+$ & 0.008 & 
$0.032 \pm 0.008 $ \\
 ~~~~~~~~~  & E2 & $14_1^+ \rightarrow 12_2^+$ & 3.4 & 
$3.6 \pm 0.5 $ \\			      
 ~~~~~~~~~  & E2 & $17_1^- \rightarrow 15_1^-$ & 2.9   & 
$3.0 \pm 1.6 $ \\			      
 ~~~~~~~~~  & E3 & $17_1^- \rightarrow  14_1^+$ & 6     & 
$ 16 \pm 6 $ \\
\end{tabular}
\end{table}


\begin{references}
\bibitem{sigma} We define 
$\sigma =\{(1/N_{d}) \sum_{i}[E_{exp}(i)-E_{calc}(i)]^{2}\}^{1/2}$, where
$N_{d}$ is the number of data.
\end{references}
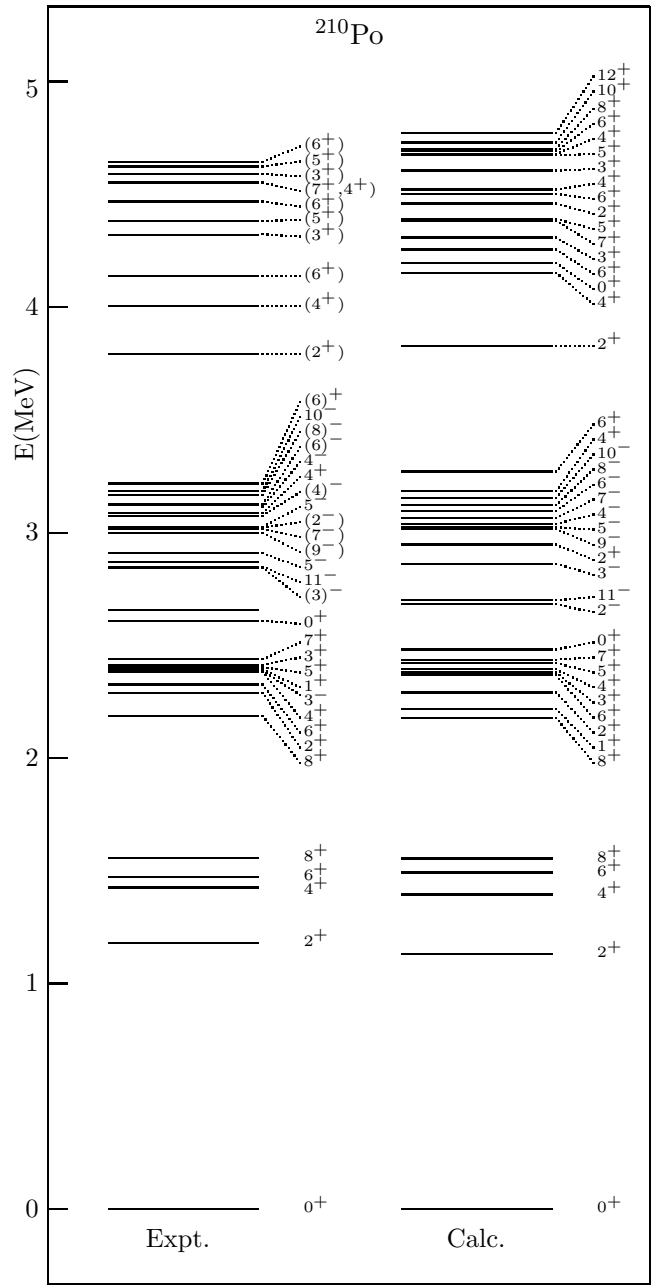
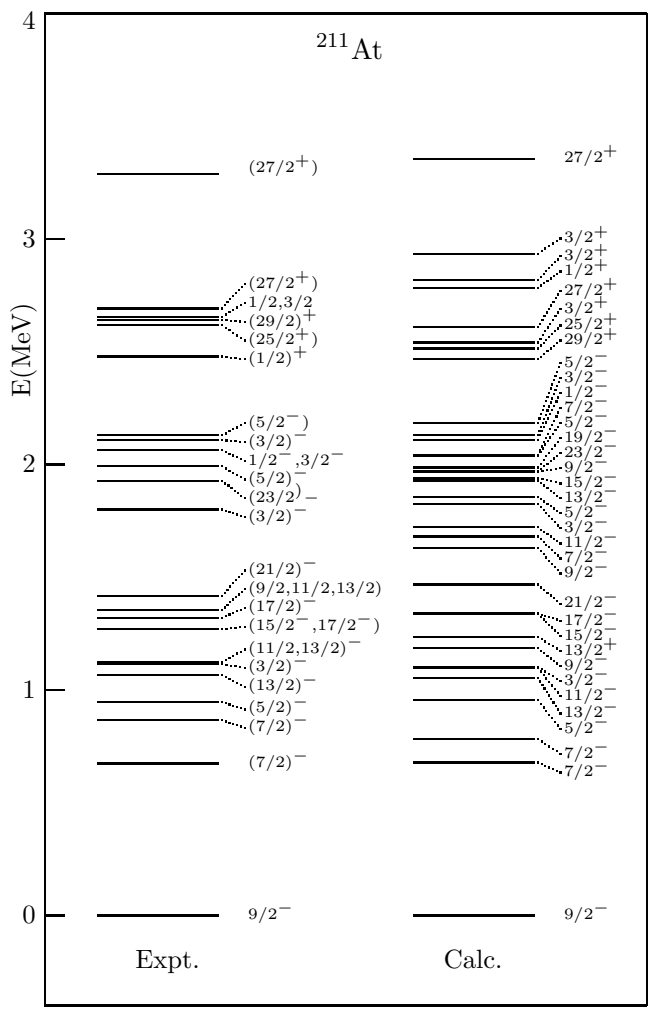
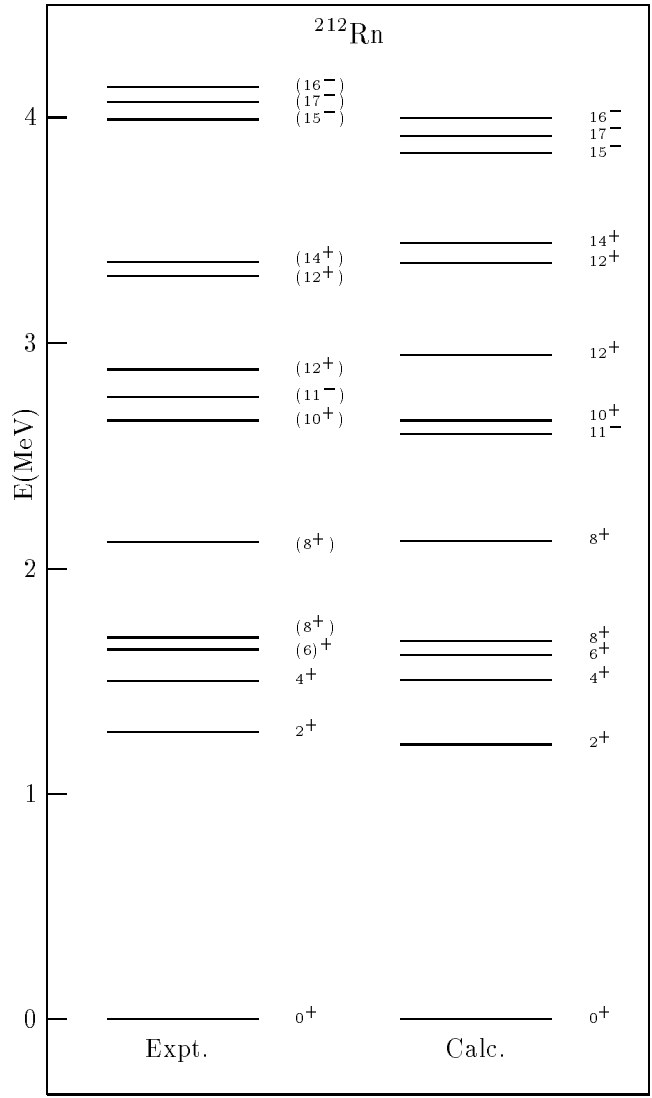
\end{document}